\documentclass[floatfix, a4paper, amsfonts, amssymb, amsmath, reprint, showkeys, nofootinbib, twoside, superscriptaddress, aps, prb]{revtex4-2}

\usepackage{graphicx}
\usepackage{physics}
\usepackage{mathtools}
\usepackage{colonequals}
\usepackage[caption=false]{subfig}
\usepackage{nicefrac}
\usepackage{color, colortbl}
\usepackage{xcolor, colortbl}

\usepackage{multirow}
\usepackage[normalem]{ulem} 
\usepackage{float}
\usepackage{xr}
\usepackage{hyperref}

\graphicspath{{Images/}}

\newcommand{\pr}{\prime}

\newcommand{\eps}{\varepsilon}
\newcommand{\mr}[1]{\mathrm {#1}}
\newcommand{\Vsup}{V_\mathrm{S}}

\newcommand{\ie}{\textit{i.e.}}
\newcommand{\eg}{\textit{e.g.}}

\renewcommand{\dd}{\mathrm{d}}

\newcommand{\eq}[1]{(\ref{#1})}

\newcommand{\ii}{\mathrm{i}}
\newcommand{\rr}{{\vb r}}
\newcommand{\kk}{{\vb k}}
\newcommand{\eqdef}{\coloneqq}
\renewcommand{\ol}{\overline}

\usepackage[T1]{fontenc}
\usepackage{notes2bib}
\bibnotesetup{
  note-name = ,
  use-sort-key = false
}
\begin{document}
\title{Symmetries and Wavefunctions of Photons Confined in 3D Photonic Band Gap Superlattices}
\author{Marek Kozo\v{n}}
\altaffiliation[Present address: ]{Pixel Photonics GmbH, Heisenbergstraße 11, 48149 Münster, Germany}
\affiliation{Complex Photonic Systems (COPS), MESA+ Institute for Nanotechnology, University of Twente, P.O. Box 217, 7500 AE Enschede, The Netherlands}
\affiliation{Mathematics of Computational Science (MACS),  MESA+ Institute for Nanotechnology, University of Twente, P.O. Box 217, 7500 AE Enschede, The Netherlands}

\author{Ad Lagendijk}
\affiliation{Complex Photonic Systems (COPS), MESA+ Institute for Nanotechnology, University of Twente, P.O. Box 217, 7500 AE Enschede, The Netherlands}

\author{Matthias Schlottbom}
\affiliation{Mathematics of Computational Science (MACS),  MESA+ Institute for Nanotechnology, University of Twente, P.O. Box 217, 7500 AE Enschede, The Netherlands}

\author{Jaap J. W. van der Vegt}
\affiliation{Mathematics of Computational Science (MACS),  MESA+ Institute for Nanotechnology, University of Twente, P.O. Box 217, 7500 AE Enschede, The Netherlands}

\author{Willem L. Vos}
\email[Corresponding author:]{w.l.vos@utwente.nl}
\affiliation{Complex Photonic Systems (COPS), MESA+ Institute for Nanotechnology, University of Twente, P.O. Box 217, 7500 AE Enschede, The Netherlands}


\begin{abstract}
We perform a computational study of confined photonic states that appear in a three-dimensional (3D) superlattice of coupled cavities, resulting from a superstructure of intentional defects. 
The states are isolated from the vacuum by a 3D photonic band gap, using a diamond-like inverse woodpile crystal structure, and exhibit 'Cartesian' hopping of photons in high-symmetry directions.
We investigate the confinement dimensionality to verify which states are fully 3D confined, using a recently developed scaling theory to analyze the influence of the structural parameters of the 3D crystal. 
We create confinement maps that trace the frequencies of 3D confined bands for select combinations of key structural parameters, namely the pore radii of the underlying regular crystal and of the defect pores. 
We find that a certain minimum difference between the regular and defect pore radii is necessary for 3D confined bands to appear, and that an increasing difference between the defect pore radii from the regular radii supports more 3D confined bands. 
In our analysis we find that their symmetries and spatial distributions are more varied than electronic orbitals known from solid state physics. 
We surmise that this difference occurs since the confined photonic orbitals derive from global Bloch states governed by the underlying superlattice structure, whereas single-atom orbitals are localized. 
We also discover pairs of degenerate 3D confined bands with p-like orbital shapes and mirror symmetries matching the symmetry of the superlattice. 
Finally, we investigate the enhancement of the local density of optical states (LDOS) for cavity quantum electrodynamics (cQED) applications.
We find that donor-like superlattices, \ie , where the defect pores are smaller than the regular pores, provide greater enhancement in the air region than acceptor-like structures with larger defect pores, and thus offer better prospects for doping with quantum dots and ultimately for 3D networks of single photons steered across strongly-coupled cavities. 
\end{abstract}
\maketitle

\section{Introduction}
\label{sec:introduction}
The confinement of light is a prominent goal of nanophotonics that is traditionally achieved via a single resonator storing light for a given time duration before it leaks away to the surrounding vacuum~\cite{Vahala2003Nature,Novotny2006book,Zhong2022Nanophotonics}. 
All over the world, a large variety of resonator structures has been realized including micropillars~\cite{Gerard1998Phys.Rev.Lett., Reithmaier2004Nature}, microdisks~\cite{Peter2005Phys.Rev.Lett.,Wang2018Opt.Lett.}, rings~\cite{Armani2003Nature,Dutt2020Science}, plasmonic resonators~\cite{Miyazaki2006Phys.Rev.Lett., Chikkaraddy2016Nature} and defects in a two-dimensional photonic crystal~\cite{Akahane2003Nature,Song2005NatureMater,Nakadai2020Appl.Phys.Express}. 
Light confinement in a single cavity has been practically utilized in applications ranging from sensing~\cite{Krioukov2002Opt.Lett.,Yasuda2021Opt.Express} and enhancing absorption~\cite{Devashish2019Phys.Rev.B}, to slowing down or trapping of photons~\cite{Baba2008Nat.Photonics, Noda2000Nature}, and to enhancing spontaneous emission~\cite{Gerard1998Phys.Rev.Lett.,Kuruma2021Appl.Phys.Lett.} and other cavity quantum electrodynamic (cQED) phenomena~\cite{Michler2003, Reithmaier2004Nature, Yoshie2004Nature, Peter2005Phys.Rev.Lett., Lodahl2015Rev.Mod.Phys.,Evans2018Science}. 

\begin{figure}
	\centering
	\includegraphics[width=\linewidth]{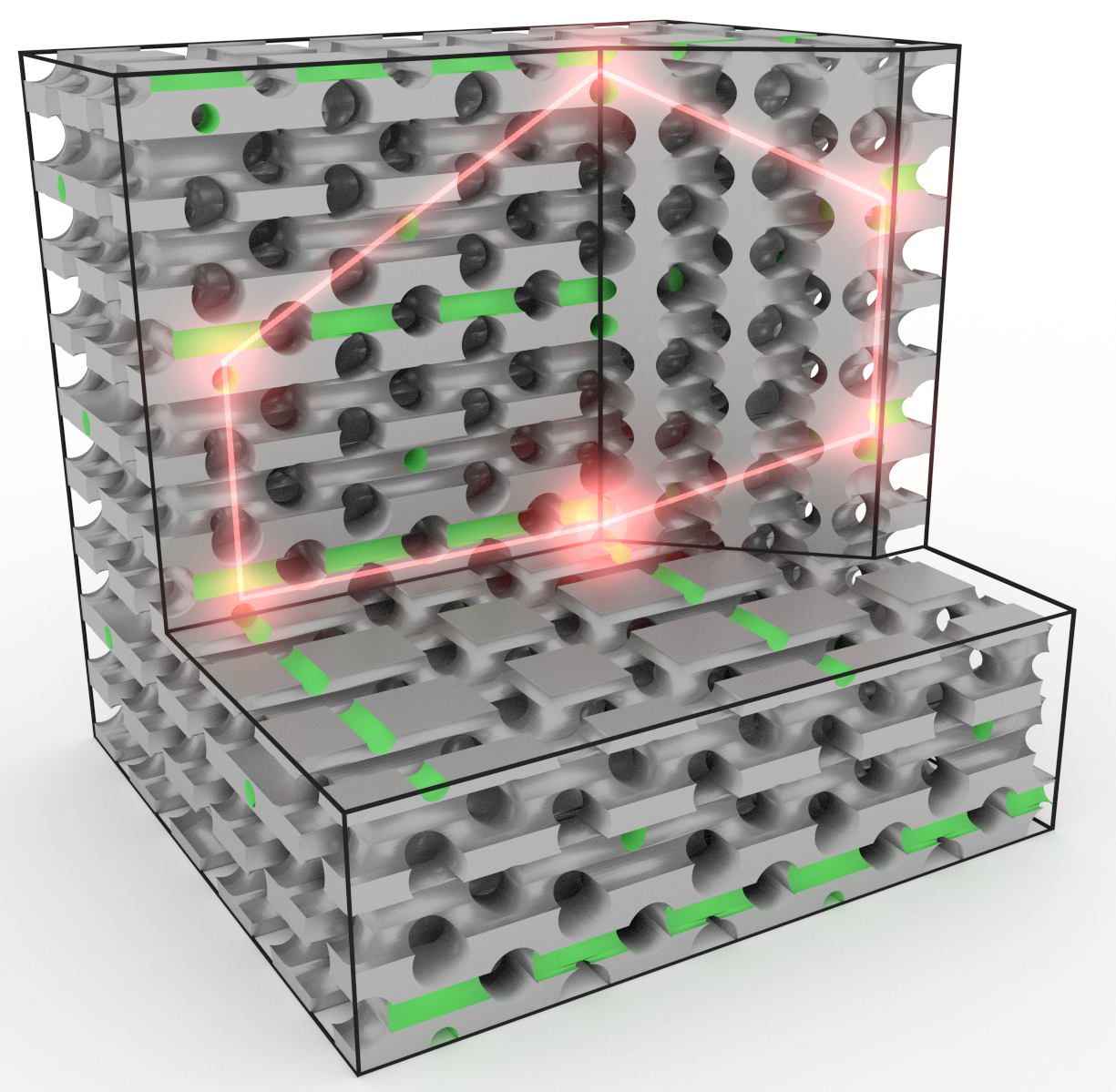}
	\caption{3D superlattice of defect pores embedded in a 3D inverse woodpile photonic band gap crystal. 
    For certain states, confined light is conceived to hop between neighboring cavities in Cartesian directions, giving rise to 'Cartesian' light. }
\label{fig:superstruc}
\end{figure}

Novel physical opportunities arise when \textit{multiple coupled} cavities are embedded in three-dimensional (3D) photonic band gap crystals as these are capable of confining light in {\it all three dimensions} simultaneously~\cite{Hack2019Phys.Rev.B, Kozon2022Phys.Rev.Lett., Adhikary2023}. 
In a perfect photonic crystal structure, thanks to multiple wave interference~\cite{vanDriel2000Phys.Rev.B} the periodic translational symmetry gives rise to a 3D \textit{photonic bandgap}~\cite{Yablonovitch1987Phys.Rev.Lett., John1987Phys.Rev.Lett., Joannopoulos2008book, VonFreymann2010AdvFunctMaterials}, that is, a range of frequencies for which light is forbidden to propagate inside the crystal irrespective of its wave vector and polarization. 
The introduction of intentional defects on a lattice superperiodic over the crystal lattice disrupts the local symmetry of the crystal, resulting in the appearance of a variety of localized states inside the band gap~\cite{Hack2019Phys.Rev.B, Kozon2022Phys.Rev.Lett.}. 
Some of these states give rise to so-called 'Cartesian light', whereby photons confined in one cavity in all three directions simultaneously, hop to a nearby cavity, as described by the well-known tight-binding approximation~\cite{Ashcroft1976book}, see Figure~\ref{fig:superstruc}. 
As a result, defect bands arise, analogous to coupled atomic orbitals that may hybridize in an atomic superlattice to semiconductor defect bands that are pursued for solar photovoltaics~\cite{Luque1997Phys.Rev.Lett.}.

The bandwidth of the photonic band gap is a crucial factor in order to effectively shield the 3D superlattice states from the surrounding vacuum states~\cite{VosLightLocalisationandLasing}, and to make the photonic cavity superlattice robust with respect to unavoidable fabrication disorder~\cite{Woldering2009J.Appl.Phys.}. 
Exceptionally wide band gaps are known to occur in crystals with diamond-like symmetry~\cite{Maldovan2004Nat.Mater.}. 
An example of these are the so-called inverse-woodpile photonic crystals, consisting of two 2D arrays of nanopores~\cite{Ho1994SolidStateCommun.}.
These structures have been realized using various nanofabrication techniques and high-index backbones~\cite{Schilling2004Appl.Phys.Lett.}. 
In the Complex Photonic Systems (COPS) chair at the University of Twente, we have developed CMOS-compatible nanofabrication methods to fabricate such crystals by etching deep pores into silicon~\cite{Woldering2008Nanotechnology,  
Tjerkstra2011J.Vac.Sci.Technol.B, Broek2012Adv.Funct.Mater., Grishina2015Nanotechnology, Goodwin2023Nanotechnology}. 
The cavity in these crystals is realized by altering the radius of two proximate orthogonal \textit{defect} pores, thereby creating an excess of one type of material in their proximal region~\cite{Woldering2014Phys.Rev.B}. 

Inverse woodpile photonic crystals with and without defects have been investigated both theoretically~\cite{Hillebrand2003J.Appl.Phys., Woldering2009J.Appl.Phys., Woldering2014Phys.Rev.B, Devashish2017Phys.Rev.B, Devashish2019Phys.Rev.B, Hack2019Phys.Rev.B} and experimentally~\cite{Schilling2004Appl.Phys.Lett., Leistikow2011Phys.Rev.Lett., Huisman2011Phys.Rev.B, Adhikary2020Opt.Express, Uppu2021Phys.Rev.Lett., Adhikary2023}. 
3D confined states in these structures have great potential for, \eg , cavity QED and Anderson localization of light.
To properly interpret the experimental data and provide guidance for fabrication, it is crucial to understand the dependence of the crystal properties on its structural parameters, namely the regular and defect pore sizes. 
Therefore, in this paper, we perform a thorough computational characterization of light confined in photonic crystals with respect to both pore sizes.
We create so-called confinement maps by keeping one of the radii constant and varying the other one. 
Simultaneously, we analyze the symmetry of selected structures and investigate the enhancement of the local density of states (LDOS)~\cite{Sprik1996Europhys.Lett.} that is crucial for potential applications in cQED~\cite{Lodahl2015Rev.Mod.Phys.}. 

\section{Methodology}
\label{sec:methodology}
\begin{figure}
	\centering
	\includegraphics[width=\linewidth]{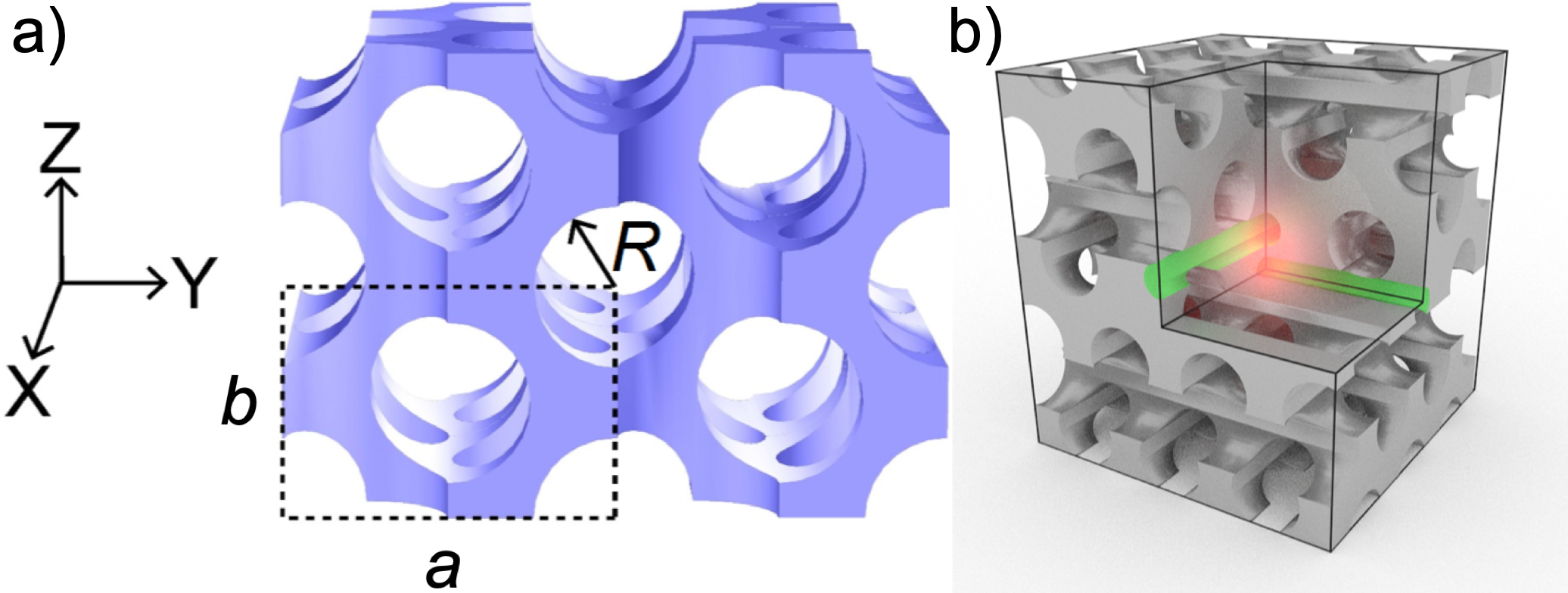}
	\caption{(a) Structure of the perfect inverse woodpile photonic crystal. We use a tetragonal unit cell with lattice constants $b$ and $a$, and the pore radius is denoted by $R$. The figure shows a $2\times2\times2$ supercell. (b) Design of a cavity. The radius of two proximal defect pores (shown in green) is altered, resulting in a region with excess of either silicon or air, which behaves as a cavity confining the light (orange glow).}
\label{fig:struc}
\end{figure}

We study 3D cavity superlattices embedded in 3D inverse woodpile photonic band gap crystals. 
The inverse woodpile crystal structure consists of two perpendicular 2D arrays of nanopores with radius $R$ in a high-refractive-index medium such as silicon~\cite{Ho1994SolidStateCommun.}, as illustrated in Figure~\ref{fig:struc}(a).
In our computations, we use the relative permittivity of silicon $\eps = 12.1$~\cite{Hillebrand2003J.Appl.Phys.}.

\begin{figure}[tb]
\centering
\includegraphics[width=\linewidth]{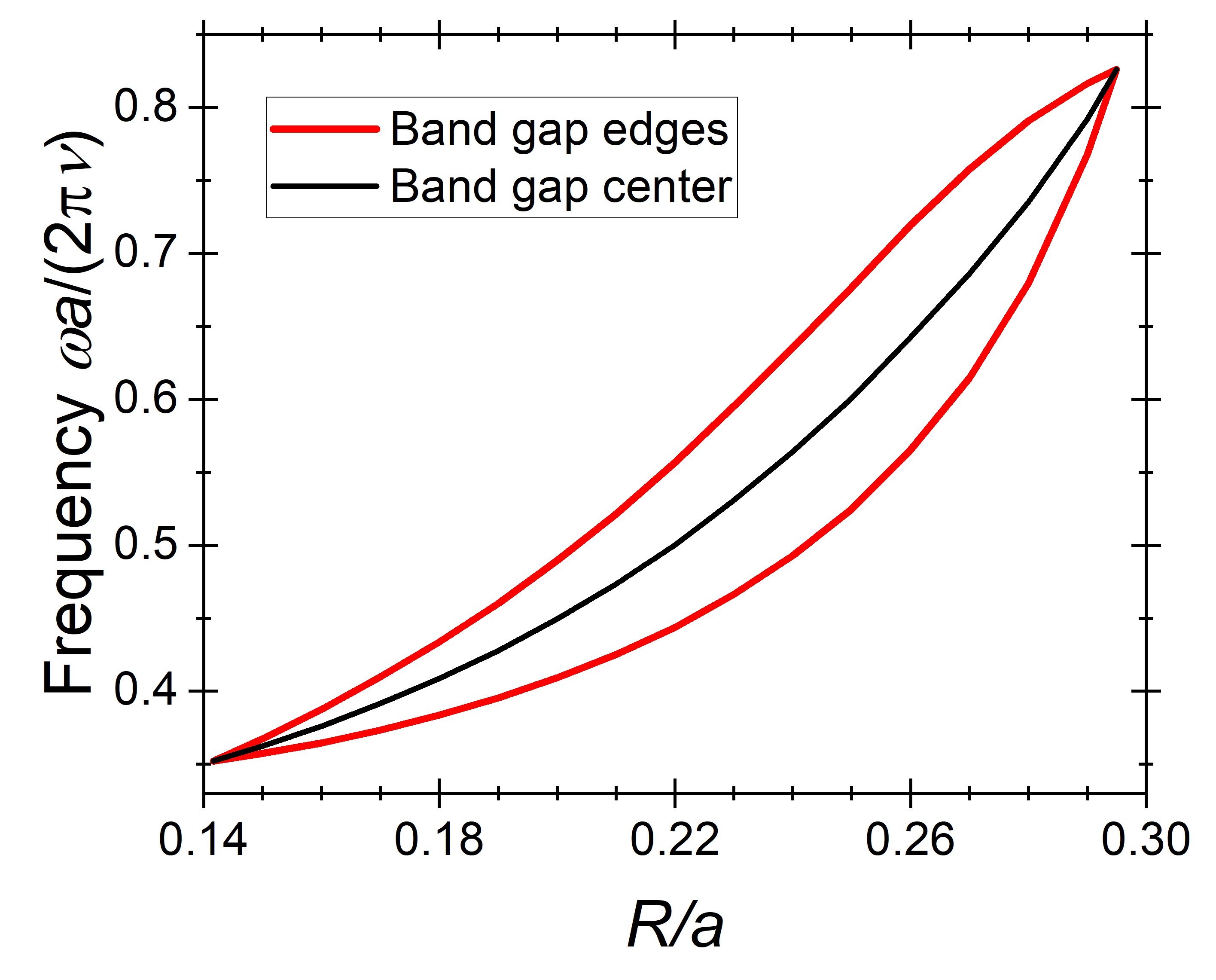}
\caption{Band gap frequencies as a function of pore radius $R$ in an inverse woodpile photonic crystal with $\eps = 12.1$ typical for a silicon backbone. 
The 3D photonic band gap exists for pore radii $0.15\le R/a\le 0.29$, with the maximum width at $R/a = 0.245$. }
\label{fig:bandgapvsR}
\end{figure}

The plane normal to each 2D pore array corresponds to the (110) crystal face of a conventional diamond structure. 
We employ a tetragonal unit cell with lattice parameters $b$ (in the $x$- and $z$- directions) and $a$ (in the $y$- direction). 
We set $a/b =\sqrt{2}$ to ensure a cubic crystal structure.  
Varying the ratio $R/a$ results in tuning of both the center frequency and the bandwidth of the band gap, as shown in Figure~\ref{fig:bandgapvsR}. 
It has been found that the widest band gap is obtained for the ideal radius $R/a=0.24$~\cite{Hillebrand2003J.Appl.Phys., Woldering2009J.Appl.Phys.}. 
Throughout this paper, we express the frequency in its reduced form $\tilde\omega=\omega a/(2\pi\nu)$, with $\nu$ the speed of light in vacuum.

We introduce a single cavity in an inverse woodpile photonic crystal by altering the radius $R'\ne R$ of two proximate perpendicular \textit{defect pores}~\cite{Woldering2014Phys.Rev.B}, as shown in Figure~\ref{fig:struc}(b).
We introduce multiple cavities by introducing defect pores at every third pore of the underlying inverse woodpile structure, giving rise to a defect superlattice of linear size $N = 3$ that is commensurate with the underlying crystal. 
The introduction of the defect superlattice causes some of the bands to move into the band gap of the perfect crystal.
The states of these defect bands are then confined in various dimensions, depending on the structure of the defect. 
We denote the number of dimensions in which a band of states is confined as the \emph{confinement dimensionality} $c$.
For more discussion on defect superlattices and wave confinement dimensionalities, see Ref.~\cite{Kozon2022Phys.Rev.Lett.}.

In this paper, we aim to find point-confined ($c = 3$) bands and investigate their dependence on the structural parameters of the inverse woodpile photonic crystal.
To this end, we employ the scaling analysis of Ref.~\cite{Kozon2022Phys.Rev.Lett.}, supplemented by the MBC clustering algorithm presented in Ref.~\cite{Kozon2023Opt.Express}.
Specifically, we utilize the scaling to identify the set of confinement dimensionalities $c$ present in the structure, which is then supplemented as an input to the MBC clustering algorithm.
Note that for small supercells, the scaling analysis tends to identify several bands as plane-confined ($c = 1$), which is unphysical since our superlattice does not contain plane defects. 
We thus automatically exclude the $c=1$ confinement dimensionality from the input into the MBC algorithm. 
We note that, even though the power of this analysis method exceeds any other known method of confinement classification, it is known to be not fully accurate for small supercells~\cite{Kozon2022Phys.Rev.Lett.}, so a few out of many bands may end up being misidentified.

Our confinement analysis requires knowledge of the energy-density distribution $W(\rr)$ in the superlattice.
We have calculated the energy densities as functions of the crystal pore radius $R/a$ and the defect pore radius $R' /R$ using the plane-wave expansion method implemented in the MPB code~\cite{Johnson2001Opt.Express}.
We normalize the density for each band so that $\int_{\Vsup} W \dd V=1$.
The band structure of the crystal with $R/a=0.24$ and $R' =0.5R$ in Figure~\ref{fig:bandstrsymmetries} has also been computed using the MPB code.
In the isosurface plots, we always chose to plot the isosurface corresponding to the third of the maximum of the energy density $W(\rr)=\max_{\Vsup}W(\rr)/3$ as a representative gauge.
For applications in spontaneous emission control, we also investigate the maximum energy density of each band, defining $\Omega\eqdef\max_{\Vsup}W(\rr)$.
High $\Omega$ then corresponds to high concentration of energy in the cavity.

\section{Photonic states beyond quasiatomic orbitals}
\label{sec:orbitals}
\begin{figure}
\centering
\includegraphics[width=\linewidth]{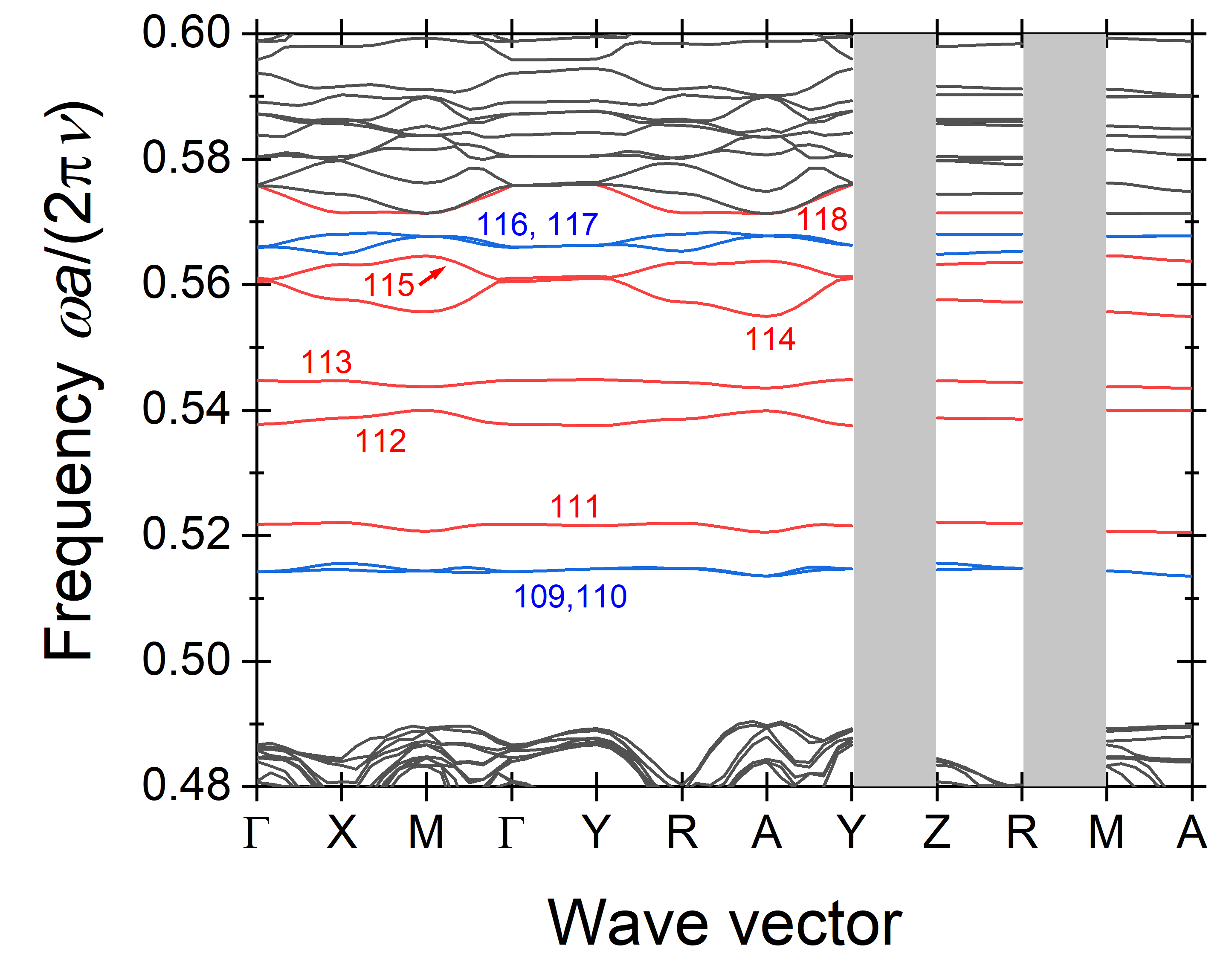}
\caption{Band structure of an inverse woodpile crystal with regular pore radius $R=0.24a$,  and the defect pore radius $R^\pr=0.5R$.
Bands that are identified as confined in $c = 3$ dimensions are colored, with red designating individual bands and blue pairs of degenerate bands.
The bands are also labeled by their band number $N_\mathrm{b}$.}
\label{fig:bandstrsymmetries}
\end{figure}

Here, we investigate the spatial energy-density profiles of several salient confined bands in the structure with $R=0.24a$, $R^\pr=0.5R$. 
As we will see, the main results are readily generalized to photonic superlattices in general. 
Some useful properties, such as band degeneracies, are best observed from the photonic band structure.
Figure~\ref{fig:bandstrsymmetries} shows the band structure of an inverse woodpile crystal with $R = 0.24a$, $R^\pr = 0.5R$.
The bands that are identified to have $c = 3$ confinement (see also Figure~\ref{fig:confmapr24rpVAR} below), have been colored to be easily recognizable.
For convenience, we distinguish the bands by their band number $N_\mathrm{b} $ that is assigned in increasing frequency order. 

First, we analyze the energy-density profile of band $N_\mathrm{b} = 111$, which has been investigated in detail by Refs.~\cite{Woldering2014Phys.Rev.B,Devashish2019Phys.Rev.B,Hack2019Phys.Rev.B}.
\begin{figure}
\centering
\includegraphics[width=0.9\linewidth]{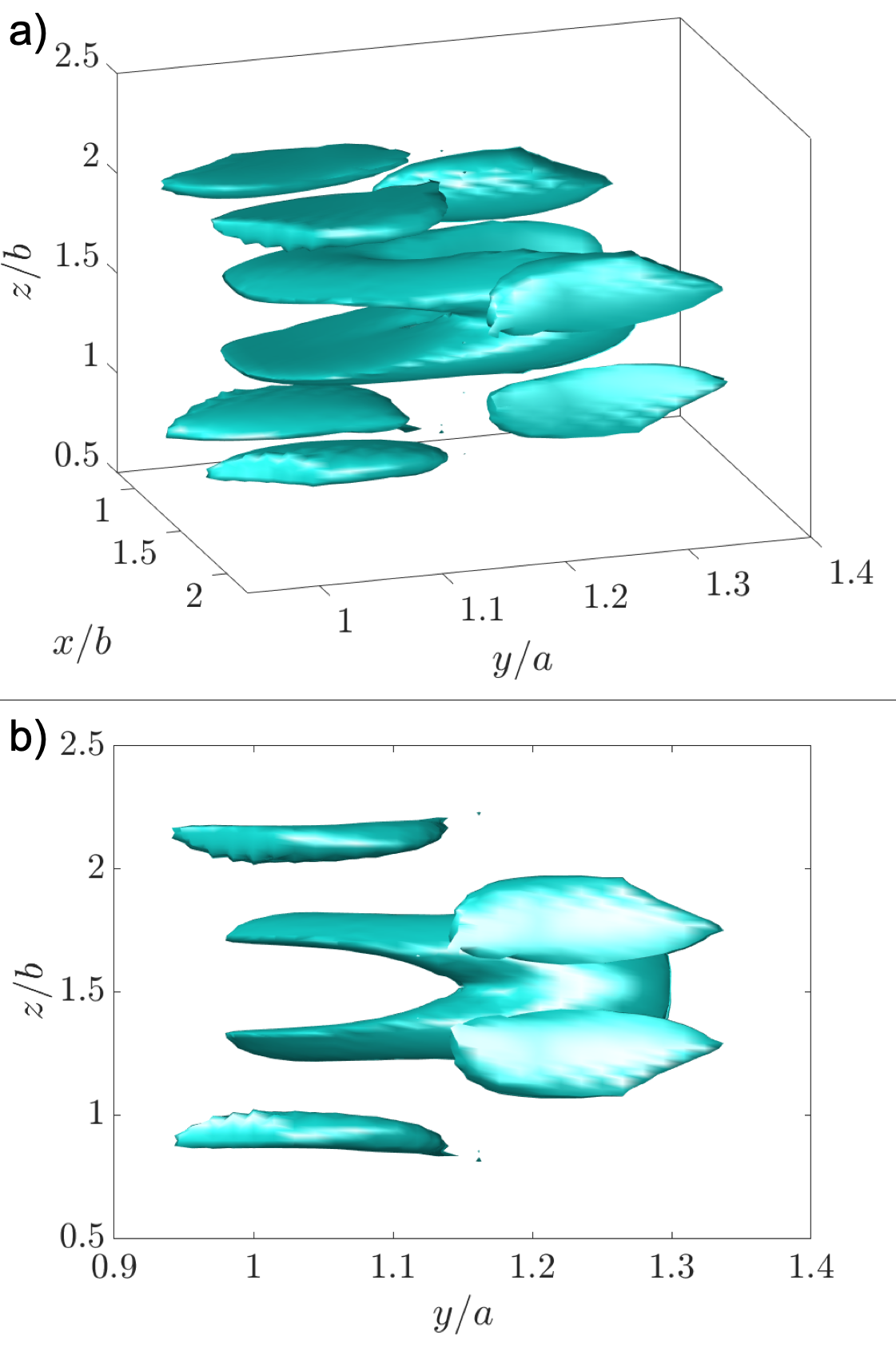}
\caption{
3D isosurface plot of the energy density in confined band $N_\mathrm{b} = 111$ in an inverse woodpile crystal with regular pore radius $R = 0.24a$, and defect pore radius $R^\pr=0.5R$. 
The energy profile exhibits specific symmetries inherited from the parent defect superlattice. 
Besides the mirror symmetries along the $z/b\approx1.5$ and $x/b\approx1.5$ planes, it is also symmetric with respect to mirroring with according to the $y/a\approx 1.15$ plane combined with 90$^{\circ}$ rotation about the $(x/b,z/b)\approx(1.5,1.5)$ axis. 
a) Birds-eye view; b) view of the $y$-$z$ plane.
}
\label{fig:band111}
\end{figure}
Specifically, Devashish \textit{et al}., based also on data of Ref.~\cite{Woldering2014Phys.Rev.B}, concluded that this band represents a quadrupole, analogous to a 3d electronic orbital~\cite{Devashish2019Phys.Rev.B}.
Here we discuss that the analogy between inverse woodpile photonic crystals and atomic orbitals is misleading and that the inverse woodpile photonic structure in fact presents a new challenge in symmetry description. 

Figure~\ref{fig:band111}(a) depicts a 3D view of the energy-density profile of the $N_\mathrm{b} = 111$ band, confined within the cavity created by the crossing defect pores. 
One can immediately see that the symmetries here are highly influenced by the defect-pore symmetry in Figure~\ref{fig:struc}(b).
There is a high-energy-density volume centered around $x/b = 1.5, y/a = 1.15, z/b = 1.5$. 
The volume is divided by the plane $y/a \approx 1.15$ into two half-spaces, where it contains a dent in each of these half-spaces.
For $y/a < 1.15$, the dent is along the $x$ direction and is surrounded in the same half-space by two smaller regions at both $z/b < 1.5$ and $z/b > 1.5$. 
For $y/a > 1.15$, the second dent in the central volume spreads in the $z$ direction and is surrounded by two smaller regions at both $x/b < 1.5$ and $x/b > 1.5$.
The energy-density profile exhibits mirror symmetries along the $z/b\approx1.5$ and $x/b\approx1.5$ plane, but not along any plane of constant $y/a$, similarly to the parent superlattice.

From the view in Figure~\ref{fig:band111}(a), it is clear that the energy density does not have the quadrupolar profile as in a 3d electronic orbital, since it lacks the $90^{\circ}$ rotational symmetry required for the 3d electronic orbital~\cite{Ashcroft1976book}. 
Moreover, we argue that the actual band symmetry is even more interesting than that of an atomic 3d orbital. 
Figure~\ref{fig:band111}(b) depicts the $y$-$z$ plane view of the energy density for the same $N_\mathrm{b} = 111$ band. 
From this view, it becomes clear that the confined band actually exhibits mirror symmetry with respect to the plane $y/a\approx1.15$ combined with a 90$^{\circ}$ rotation about the $(x/b,z/b)\approx(1.5,1.5)$ axis.
We attribute this effect to the fact that the photonic orbitals derive from global Bloch states governed by the underlying superlattice structure, whereas electronic orbitals are localized.

This discovery brings us to an extremely interesting fundamental question: 
Instead of striving for strict analogies in symmetries between the electronic orbitals and photonic bands, photonic structures could be utilized to create photonic analogies of ``orbitals'' with a much greater variety of geometries and symmetries than feasible in spherical atoms~\cite{Sakurai1994}, and controllable in more complicated electronic systems. 
Since there is a great diversity of photonic crystals of various structures and a plethora of options to introduce defects of different kinds in them (see, \eg , Refs.~\cite{Maldovan2004Nat.Mater.,Joannopoulos2008book}), it seems highly plausible that the set of symmetries achievable by ``photonic orbitals'' could be much more numerous and varied than that occurring in electronic orbitals. 
Since it is well known from atomic solid-state physics that the symmetries of atomic orbitals are closely tied to the appearance of the band structure and even to the macroscopic behavior of materials, such as their placement in the periodic table~\cite{Ashcroft1976book}, it is exciting to investigate to what extent the symmetries in ``photonic orbitals'' translate to the macroscopic behavior of ``photonic solid-state matter'' and how their great variety can be utilized. 

\section{Symmetry and degeneracy}
\label{sec:symmetry-degeneracy}
\begin{figure}
\centering
\includegraphics[width=\linewidth]{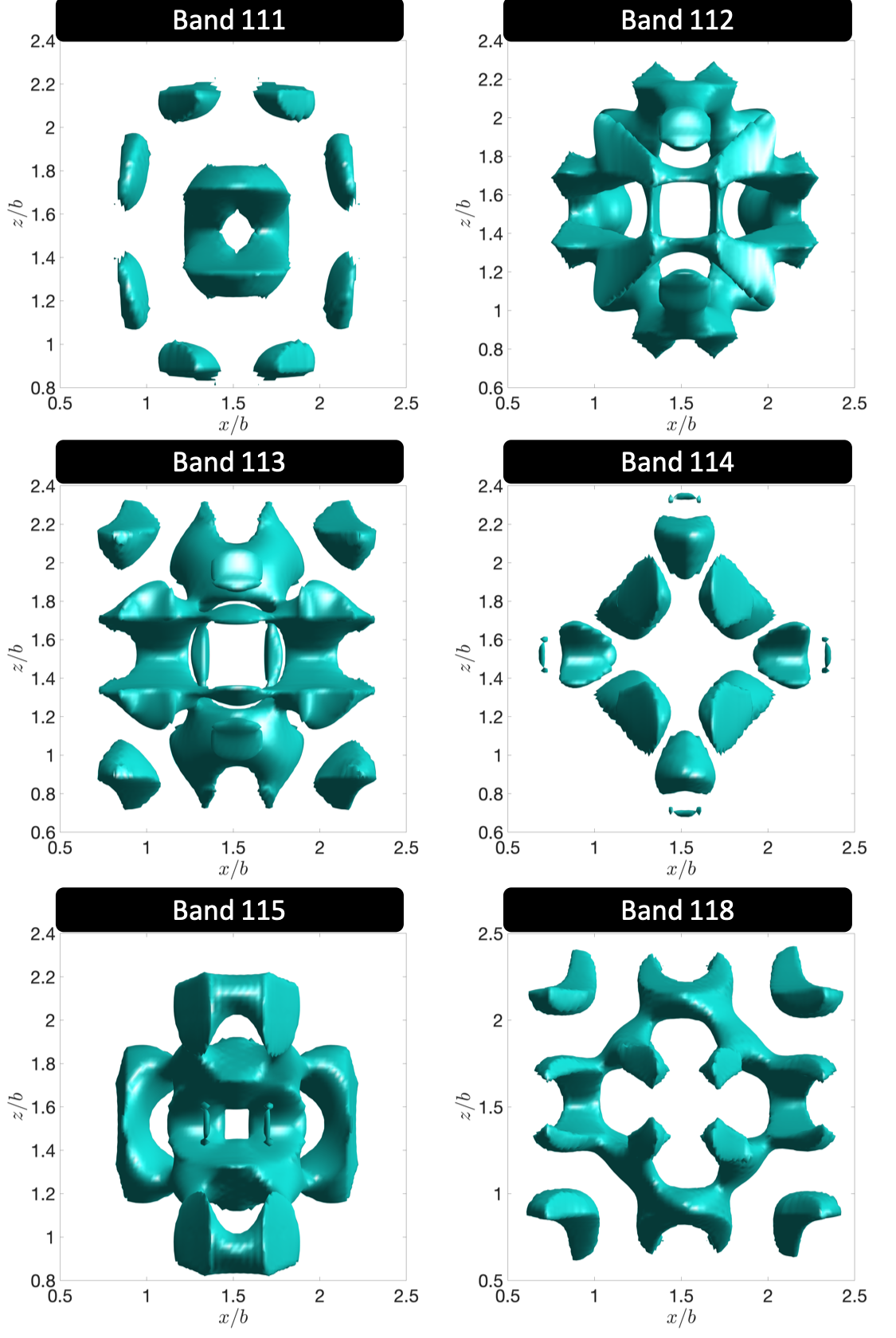}
\caption{
The $x$-$z$ plane profile of the energy-density distribution of the nondegenerate 3D confined bands in an inverse woodpile crystal with the regular pore radius $R = 0.24a$, and the defect pore radius $R^\pr = 0.5R$.
The plane view may give an impression of the 90$^{\circ}$ rotational symmetry of these profiles, but that is misleading, as is easily verified by looking at the 3D profile of the $N_\mathrm{b} = 111$ band in Figure~\ref{fig:band111}.
}
\label{fig:3Dindepbands}
\end{figure}

Figure~\ref{fig:3Dindepbands} depicts the $x$-$z$ view of all nondegenerate 3D confined bands in Figure~\ref{fig:bandstrsymmetries}. 
All bands depicted appear to have unrelated spatial energy-density profiles, which agrees with the fact that these bands are nondegenerate. 
Even though they appear to have $90^{\circ}$ rotational symmetries, this is an optical illusion created by the plane-view of the plots, as is readily seen by inspecting different views of the $N_\mathrm{b} = 111$ band in Figs.~\ref{fig:3Dindepbands} and~\ref{fig:band111}. 
Nevertheless, it is important to stress that these profiles exhibit mirror symmetries with respect to the $x/b = 1.5$ and $z/b = 1.5$ planes, hence passing through the axes of each defect pore, respectively.
Note that it also appears that the $N_\mathrm{b} = 118$ band is degenerate with the $N_\mathrm{b} =119$ band, which has not been identified as confined. 
This is likely due to the decreased accuracy of the employed scaling method for small supercells, as previously described in detail by us~\cite{Kozon2022Phys.Rev.Lett.,Kozon2023Opt.Express}.

\begin{figure}
\centering
\includegraphics[width=\linewidth]{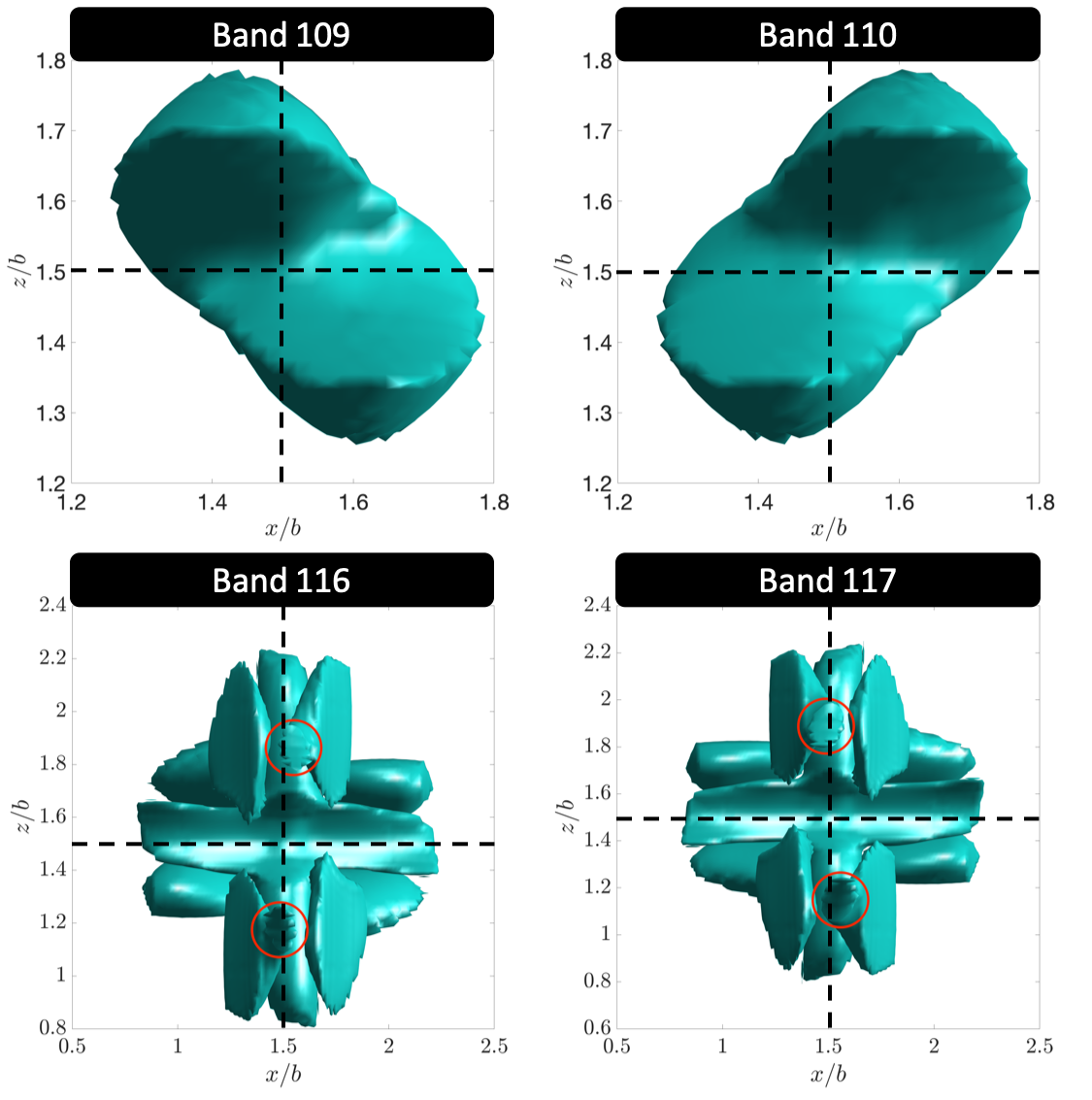}
\caption{The $x$-$z$ plane profile of the energy-density distribution of the degenerate pairs of 3D confined bands in an inverse woodpile crystal with regular pore radius $R=0.24a$, and defect pore radius $R^\pr=0.5R$.
The dashed lines indicate the planes of the mirror symmetry between these bands and the red circles for the bands $N_\mathrm{b} =116,117$ indicate the profile parts where the mirror symmetries can be easily spotted.}
\label{fig:3Ddegbands}
\end{figure}

Figure~\ref{fig:3Ddegbands} depicts the two pairs of degenerate bands $N_\mathrm{b} = 109, 110$ and $N_\mathrm{b} = 116, 117$. 
These bands are related to each other by the mirror symmetries along the planes $x/b \approx 1.5$ and $z/b \approx 1.5$.
In solid-state physics, atomic orbitals are categorized as different spherical multipoles~\cite{Ashcroft1976book}. 
This is possible due to the spherical symmetry of the atomic geometry, with each multipole exhibiting a lower symmetry than spherical symmetry. 
One thus obtains three mutually orthogonal dipoles, each exhibiting a $180^\circ$ rotational symmetry along every plane.
Whereas the inverse woodpile structure obviously does not possess spherical symmetry, it possesses lower symmetry, namely only the $x$-$z$ plane is rotationally symmetric with respect to 180$^\circ$. 
This symmetry thus only allows for two mutually orthogonal dipoles, which are both symmetric with respect to 180$^\circ$ rotation in the $x$-$z$ plane.
Exactly this property is exhibited by the band pairs $N_\mathrm{b} = 109, 110$ and $N_\mathrm{b} = 116, 117$. 
We therefore interpret these two pairs of degenerate bands as generalized dipoles, for the case of the inverse woodpile structural symmetry. 
Moreover, within the limited set of the energy-density profiles that we have visually investigated, it seems that if the energy-density profile of a band is not symmetric with respect to both the mirror planes $x/b=1.5$ and $z/b=1.5$, the band turns out to be degenerate with another band which then complements these mirror symmetries.
This symmetry relation could therefore be used to spot or confirm the presence of degenerate bands in the inverse woodpile photonic band structure and possibly even be generalized to other situations.

\begin{figure}
\centering
\includegraphics[width=\linewidth]{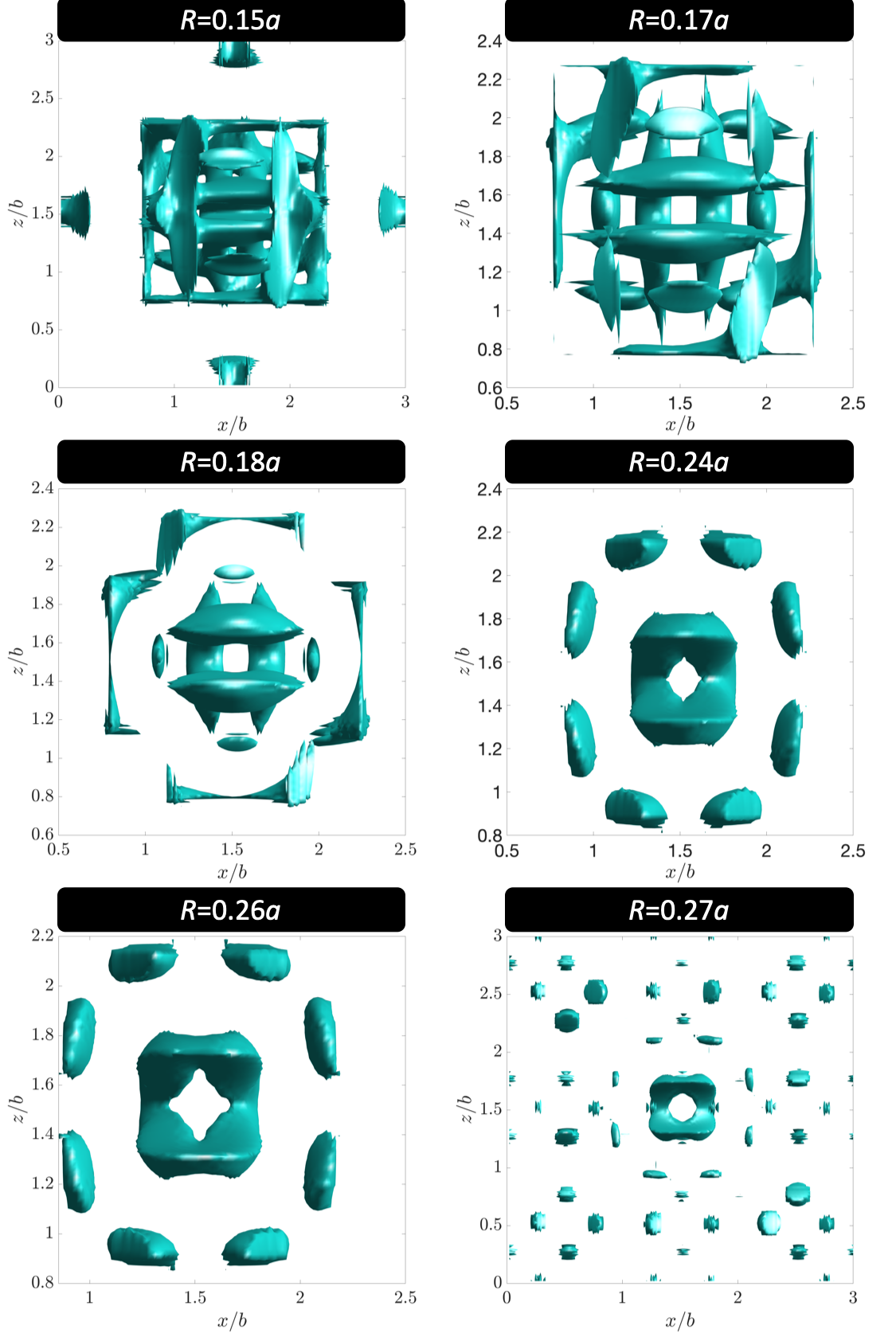}
\caption{The $x$-$z$ plane profile of the energy-density distribution of the $N_\mathrm{b} = 111$ band in an inverse woodpile crystal with varying regular pore radius $R$, and the defect pore radius $R^\pr = 0.5R$.}
\label{fig:111evol}
\end{figure}

Finally, Figure~\ref{fig:111evol} depicts the evolution of the $N_\mathrm{b} =111$ band in structures with increasing pore radii $R$, while maintaining the ratio $R^\pr/R$. 
The energy-density maximum of the band changes with increasing $R$, which makes it difficult to plot the same isosurface every time. 
Nevertheless, the mode seems to maintain similar symmetric profile shape with strongly varying pore radius. 
This visualisation confirms the results from Figure~\ref{fig:confmaprVARrp0.5relfreq} below, that for the radii $R=0.15a$ and $R=0.27a$ this band is not 3D confined, since in both cases the profile extends throughout the whole supercell, at least in the $x$-$z$ plane.
It is remarkable that even for $R=0.27a$, the shape of the central volume still resembles the confined profile seen for lower radii.
Overall, Figure~\ref{fig:111evol} offers the interesting observation that, for a given band, its energy-density profile is robust to strongly varying structural parameters of the underlying inverse woodpile structure. 

\section{Enhanced local density of states and cavity QED} 
\label{sec:enhanced}

It is well-known that in thermodynamic equilibrium the time-averaged energy density $\ol{W_\omega}$ of the electromagnetic field at frequency $\omega$ corresponds to the product of the average energy per mode $\ol w(\omega,T)$ and the local density of states (LDOS) $N(\rr,\omega)$: 
\begin{equation}
\ol{W_\omega(\rr,\omega)} = \ol w(\omega,T)N(\rr,\omega),
\label{eq:BB}
\end{equation} 
where $T$ is the temperature, see, \textit{e.g.}, Ref.~\cite{Novotny2006book}. 
Expression~\eq{eq:BB} indicates that by manipulating the energy density $W$, we control the LDOS, which is a crucial control mechanism in cQED~\cite{Novotny2006book}.
A large LDOS is favorable for cQED applications, initially for enhanced spontaneous emission and eventually, at even larger LDOS, for cQED strong coupling whereby quantum matter states are hybridized with photonic states~\cite{Reithmaier2004Nature, Yoshie2004Nature, Peter2005Phys.Rev.Lett.,Evans2018Science,Trivedi2019Phys.Rev.Lett.,Kuruma2021Appl.Phys.Lett.,Karnieli2023Sci.Adv.}. 
By positioning quantum dots within the pores of inverse woodpile photonic crystal superlattices~\cite{Schulz2022ACSNano} and coupling them at the correct electromagnetic frequencies, it will be feasible to observe these cQED phenomena. 
It is therefore important to investigate the energy-density enhancement properties of the inverse woodpile photonic crystals with respect to their structural parameters.

As discussed below in Section~\ref{sec:confmaps} in Figs.~\ref{fig:confmapr24rpVAR}-\ref{fig:confmaprVARrp0.5relfreq}, it appears that larger pore radii $R$ are generally more favorable towards confining light with large local enhancements of the optical energy density. 
Out of all investigated bands, $N_\mathrm{b} =108$ in the case of the $R=0.27a,R^\pr=1.2a$ structure has the largest maximum energy density $\Omega = 29.6$. 
We therefore investigate the energy profile of this acceptor-like band in greater detail. 

\begin{figure}
\centering
\includegraphics[width=\linewidth]{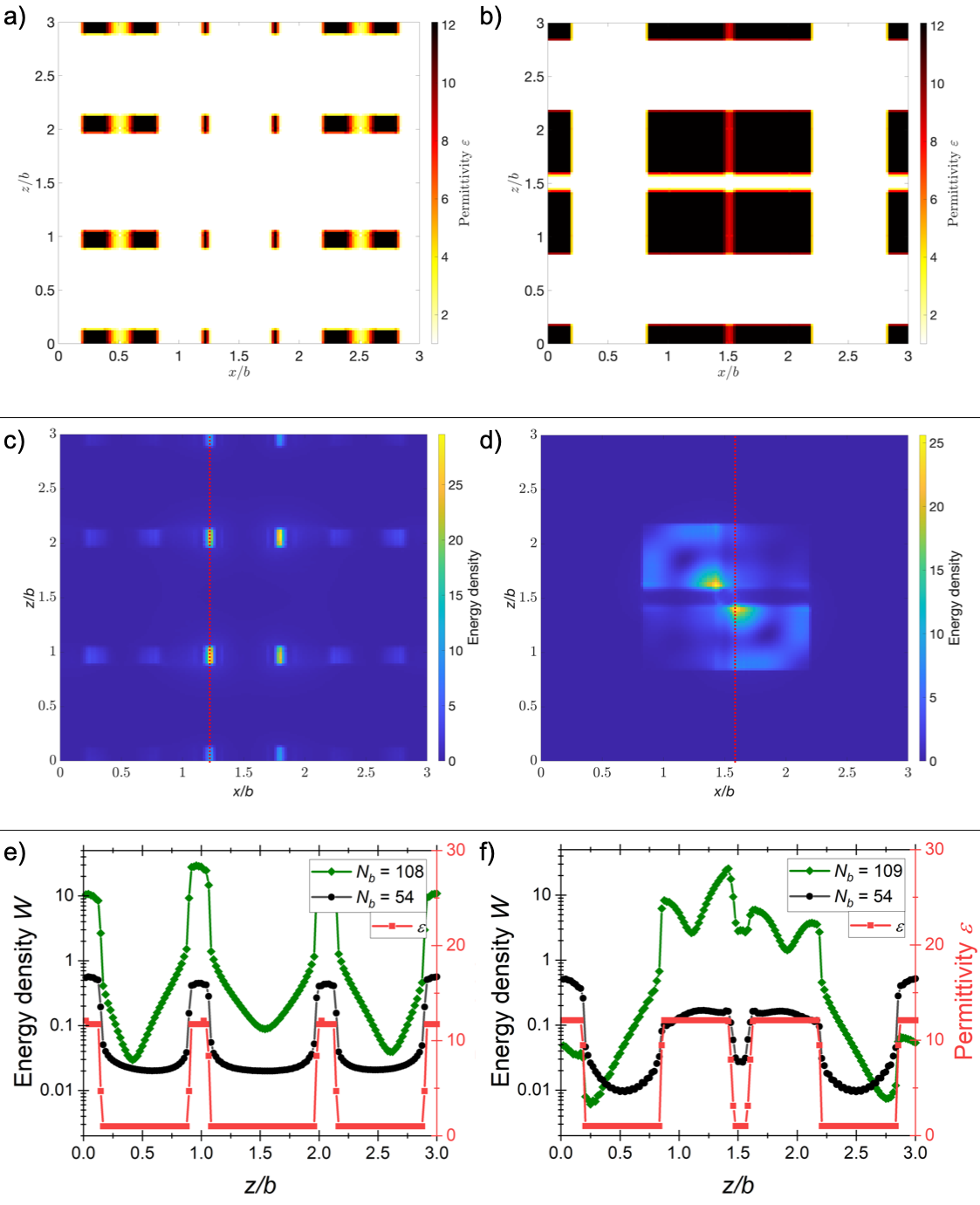}
\caption{Energy enhancement in two different inverse woodpile structures. 
(a), (c), (e) Acceptor-like structure with the regular pore radius $R=0.27a$, defect pore radius $R^\pr=1.2R$, band number $N_\mathrm{b} =108$. 
(b), (d), (f) Donor-like structure with the regular pore radius $R=0.26a$, defect pore radius $R^\pr=0.5R$, band number $N_\mathrm{b} =109$. 
(a) Permittivity distribution of the acceptor-like structure in the $y/a=1$ plane. 
(b) Permittivity distribution of the donor-like structure in the $y/a=1.13$ plane.
(c) Energy density distribution of the acceptor-like band $N_\mathrm{b} =108$ in the $y/a=1$ plane.
(d) Energy density distribution of the donor-like band $N_\mathrm{b} =109$ in the $y/a=1.13$ plane.
(e) Cross-section of the energy density distribution for the $(x/b,y/a)=(0.87,1)$ line, denoted by the red line in (c). 
Depicted is the confined band $N_\mathrm{b} =108$ and a reference extended band $N_\mathrm{b} =54$.
(f) Cross-section of the energy density distribution for the $(x/b,y/a)=(1.60,1)$ line, denoted by the red line in (d). 
Depicted is the confined band $N_\mathrm{b} =109$ and a reference extended band $N_\mathrm{b} =54$.}
\label{fig:enenh}
\end{figure}

Figure~\ref{fig:enenh}(a) shows the cross-section of the permittivity of the investigated structure in the plane $y/a=1$. 
Figure~\ref{fig:enenh}(c) then depicts the energy density distribution $W$ of the band $N_\mathrm{b} =108$.
To investigate the energy density more quantitatively, we plot $W$ along the red line in the figure, corresponding to $(x/b,y/a)=(0.87,1)$, resulting in the plot in Figure~\ref{fig:enenh}(e).
Figure~\ref{fig:enenh}(e) also contains an extended band $N_\mathrm{b} =54$ for reference, to help us establish the vacuum energy density level by looking at its value of $W$ in the air regions.
This vacuum energy density turns out to be around $W=0.03$.
From this graph, it is clear that the band $N_\mathrm{b} =108$ has large energy-density peaks but these are restricted to the regions of high permittivity.
In the central cavity region, the distribution of $W$ within the large pore varies considerably and is an order of magnitude greater at the pore walls than at the pore center.
Taking into account that quantum dots embedded in a crystal usually stick to the silicon walls, this structure could provide an LDOS enhancement of around one order of magnitude compared to the vacuum level.
Nevertheless, overall, it appears that the lack of silicon and air regions that are too large significantly restrict the energy density distribution in the acceptor-like structure, thus making it a sub-par candidate for cQED applications.

To observe the influence of donor-like structures with $R^\pr<R$ on the energy density enhancement, we investigate the band $N_\mathrm{b} =209$ in the superlattice with regular pore radius $R=0.26a$ and $R^\pr=0.5R$, which exhibits the maximum of the energy density $\Omega=25.7$ that is the largest among the donor-like structures we researched.
The permittivity distribution of this structure in the plane $y/a=1.13$ is depicted in Figure~\ref{fig:enenh}(b) and the energy density profile of the band $N_\mathrm{b} =209$ in the same plane is in Figure~\ref{fig:enenh}(d).
Figure~\ref{fig:enenh}(f) shows the energy density profile along the red line in Figure~\ref{fig:enenh}(d), given by $(x/b,y/a)=(1.60,1)$.
Again, we depict also an extended band $N_\mathrm{b} =54$ for reference.
In this case, since the air volume is much smaller compared to the amount of silicon around the cavity, the energy density within the central defect pore drops only slightly compared to its values in the surrounding silicon.
Overall, even in the center of the defect pore where $W$ is the smallest, the energy-density enhancement compared to the vacuum level at $W=0.01$, and therefore also the LDOS enhancement, is more than two orders of magnitude.

To conclude this section, we observed three competing phenomena with regard to cQED applications:
Firstly, donor-like cavities are preferable for cQED due to larger silicon and smaller air regions.
Secondly, as explained below in more detail, larger defect pore deviations, \ie , smaller defect pore radii $R^\pr$ for donor-like structures, favor the appearance of more confined bands.
Thirdly, larger defect pore radii $R^\pr$, appear to favor more concentrated energy density.
From our investigation it therefore seems that a good balance between these three requirements could be around the defect pore radius $R^\pr\approx0.6R$ slightly above half the regular pore radius $R$.

\section{Confinement maps}
\label{sec:confmaps}
\subsection{Confinement versus defect pore radius}
\begin{figure}
\centering
\includegraphics[width=\linewidth]{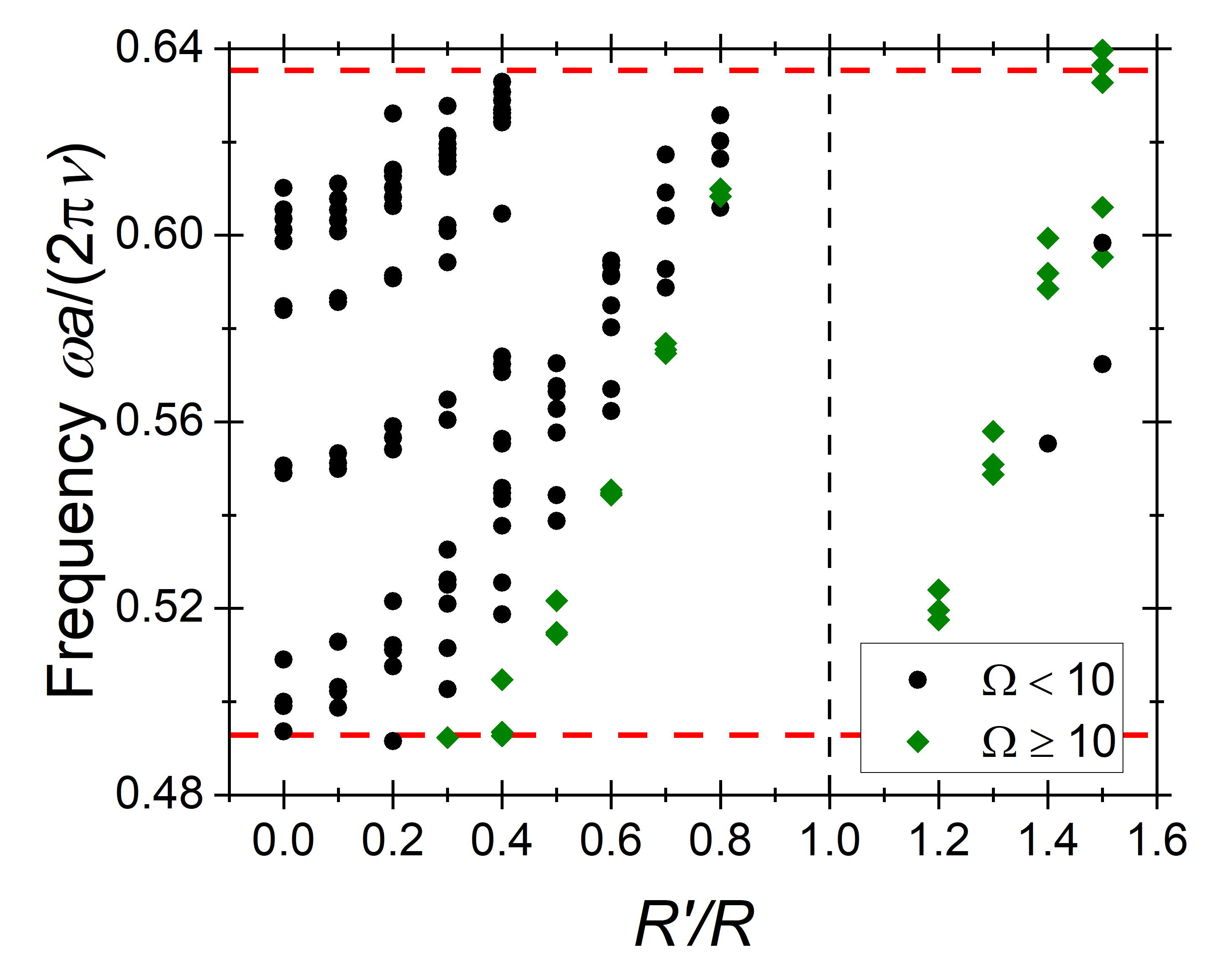}
\caption{Confinement map of point-confined $c=3$ bands in an inverse woodpile superlattice with regular pore radius $R=0.24a$ and varying radius $R^\pr$ of the defect pores. 
The perfect crystal case corresponds to $R^\pr/R=1$ and is denoted by the dashed vertical line.
Each point represents the average reduced frequency of a point-confined band and the dashed red lines represent the edges of the band gap of a perfect crystal. 
The color and shape of the symbols correspond to their maximum energy densities $\Omega$, as given in the legend.}
\label{fig:confmapr24rpVAR}
\end{figure}
Figure~\ref{fig:confmapr24rpVAR} depicts the confinement map of $c=3$ point-confined bands in an inverse woodpile superlattice with regular pore radius $R=0.24a$, while varying radius $R^\pr$ of the defect pore.
We observe four salient features.
Firstly, there is threshold in the defect pore radius, that is, a certain minimal deviation of the defect pore radius from the regular one necessary for the 3D confined bands to be present.
In this case this threshold appears to be $R^\pr\le 0.8 R$ or $R^\pr\ge 1.2 R$.

Secondly, we observe that for $R^\pr<R$, the confined bands descend into the band gap from its upper band edge, while for $R^\pr>R$, the confined bands ascend from the lower band edge.
This notion agrees with the analysis of Ref.~\cite{Villeneuve1996Phys.Rev.B} and with the semiconductor analogy, where $R^\pr<R$ corresponds to donor doping, while $R^\pr>R$ is analogous to acceptor doping~\cite{Ashcroft1976book}.
In the photonic case, we interpret this behavior using the energy functional 
\begin{equation}
\mathcal{U}(\vb E_\kk)=\frac{\int_{\Vsup} |\grad_\kk\times \vb E_\kk|^2 \dd V}{\int_{\Vsup} \eps |\vb E_\kk|^2\dd V},
\label{eq:EnfuncEE}
\end{equation} 
where $\grad_\kk\eqdef \grad+\ii\kk$, $\Vsup$ denotes the superlattice volume and $\vb E_\kk$ is the periodic part of a specific mode corresponding to the wave-vector $\kk$.
It is known that each higher-frequency mode $\vb E_\kk$ minimizes the functional $\mathcal{U}(\vb E_\kk)$ in the space orthogonal to all the lower-frequency modes, see Ref.~\cite{Joannopoulos2008book}.
The band gap appears if there is a large frequency difference between two consecutive bands. 
Decreasing the size of the defect pores $R^\pr<R$, results in additional high-index silicon in the crystal, which provides more opportunity to concentrate the electromagnetic energy in the high-index material, and thus allows for the minimum of the energy functional at lower frequency, pushing the bands from the top of the band gap downwards. 
On the other hand, increasing the size of the defect pores $R^\pr>R$ results in more air in the structure and thus restricts the freedom of concentrating the energy in the high-index material for the bands below the band gap, thereby pushing them upwards into the gap.
In this regard, it is also relevant that larger deviations of the defect pore radius $R^\pr$ from the regular pore radius $R$ provide more 3D confined bands. 

Thirdly, the confined bands exhibit a clear upward moving trend with increasing $R^\pr/R$, until they disappear in the top edge of the band gap, creating groups separated by frequency gaps.
It is worth noting that even though there are more confined bands for small defect pore sizes, only the large enough defect radii $R^\pr\ge0.3R$ provide the value of $\Omega>10$, corresponding to large energy concentration.

Fourthly and finally, these defect bands do not abruptly disappear at the edges of the band gap, but sometimes extend slightly beyond them, specifically, some confined bands cross the bottom edge of the gap for $R^\pr<R$, whereas certain bands cross the top gap edge for $R^\pr>R$.
This is because the decrease of the defect pore radius increases the total silicon volume fraction of the medium, thus effectively shifting the whole band structure slightly downwards for $R^\pr<R$, whereas the opposite happens for $R^\pr>R$, where the lower silicon volume fraction shifts the band structure slightly upwards.

We note that the specific case of $R=0.24a, R^\pr=0.5R$ has been previously investigated by means of a somewhat naive band structure analysis by Refs.~\cite{Woldering2014Phys.Rev.B,Devashish2019Phys.Rev.B,Hack2019Phys.Rev.B}, all of which found only five 3D confined bands.
Using our novel confinement identification method from Ref.~\cite{Kozon2022Phys.Rev.Lett.,Kozon2023Opt.Express}, we discovered in total 10 3D confined bands for the same physical conditions. 

\begin{figure}
\centering
\includegraphics[width=\linewidth]{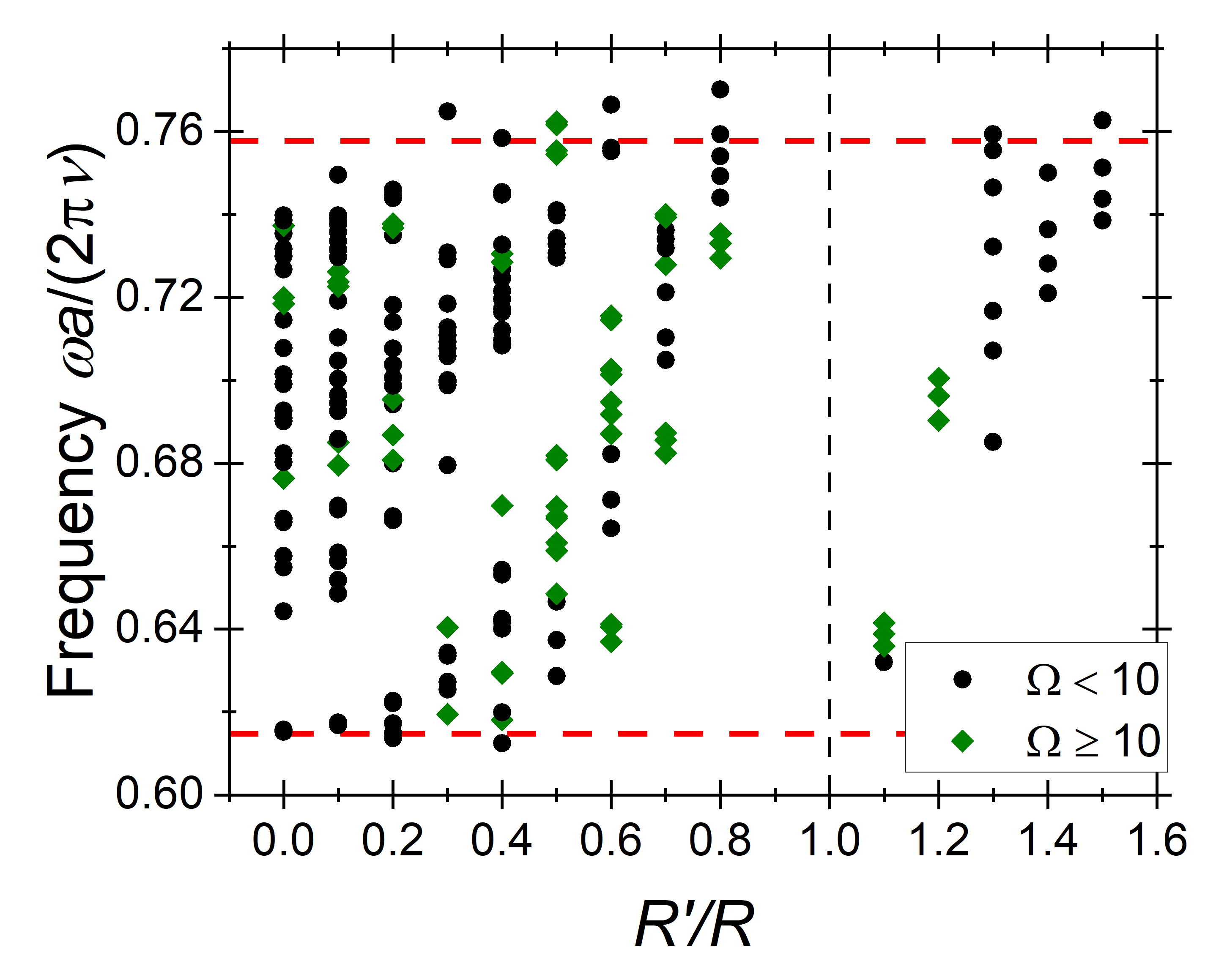}
\caption{Confinement map of point-confined $c=3$ bands in an inverse woodpile superlattice with regular pore radius $R=0.27a$ and varying radius $R^\pr$ of the crossing defect pores.
The case of the unperturbed crystal corresponds to $R^\pr/R=1$ and is denoted by the dashed vertical line.
Each point represents the average reduced frequency of a point-confined band and the dashed red lines represent the edges of the band gap of an unperturbed crystal.
The color and shape of the symbols correspond to their values of maximum energy density $\Omega$, as described in the legend.}
\label{fig:confmapr27rpVAR}
\end{figure}
Figure~\ref{fig:confmapr27rpVAR} depicts the confinement map of $c=3$ point-confined bands in an inverse woodpile superlattice with larger regular pore radius $R = 0.27a$, while varying the radius $R^\pr$ of the defect pore.
The behavior of bands here is similar to that for $R = 0.24a$ in Figure~\ref{fig:confmapr24rpVAR}, with groups of bands moving from the bottom edge of the band gap upwards and disappearing at the top edge, as $R^\pr/R$ increases. 
There are, however, three qualitative differences compared to the $R = 0.24a$ case. 
Firstly, in this case, $c=3$ defect bands are observed already for $R^\pr=1.1R$, thus reducing the pore radius threshold 
compared to $R=0.24a$. 
Secondly, the bands with high energy concentration $\Omega>10$ are now sprinkled across the whole plot, suggesting that energy concentration prefers larger pore radii $R$. 

Thirdly, we see here several bands exceeding the top of the band gap for $R^\pr<R$, which cannot be explained by the differences in silicon volume fraction.
These bands are indeed 3D confined, as confirmed by visually inspecting their energy density distribution, so this is not an artifact of the employed method. 
This observation confirms what has been already hinted at by the results of Ref.~\cite{Kozon2023Opt.Express}, namely that the confinement does not suddenly stop at the top edge of the band gap, for $R^\pr<R$. 
At frequencies above the point-confined $c=3$ bands, linearly confined $c=2$ bands appear, as seen in Ref.~\cite{Kozon2023Opt.Express}. 
It follows from this observation that the bands at the top of the band gap only seem to lose their confinement properties gradually, transitioning from $c=3$ through $c=2$ until they become extended $c=0$ bands. 
Such (partially) confined bands outside of the band gap could then possibly correspond to symmetry-protected bound states in continuum, see Ref.~\cite{Hsu2016NatRevMater}.
In contrast, the bottom edge appears to be a much stricter boundary, even when the slight shift with respect to the change in the silicon volume fraction is accounted for. 
This has been also observed in experiments, where the position of this edge provided great help when analyzing the wave confinement and connecting the experiments with the theory~\cite{Adhikary2023}. 
Analogously, the role of the band edges is exchanged, and for $R^\pr>R$, the top edge of the band gap acts as a hard boundary, only slightly shifted by the change in the silicon volume fraction, while at the bottom edge, the bands seem to lose their confinement properties only gradually.

\begin{figure}
\centering
\includegraphics[width=\linewidth]{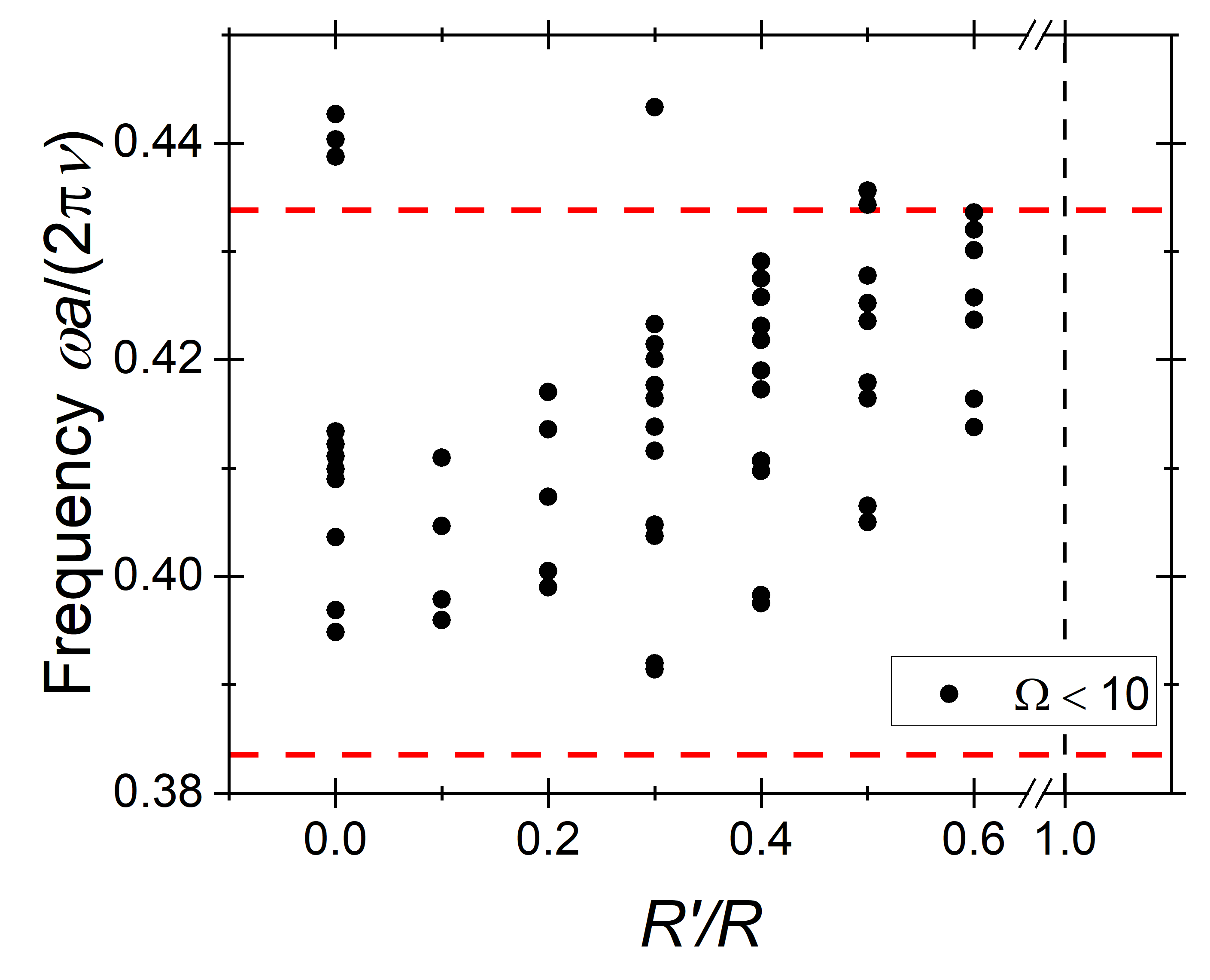}
\caption{Confinement map of point-confined $c=3$ bands in an inverse woodpile superlattice with regular pore radius $R=0.18a$ and varying radius $R^\pr$ of the crossing defect pores. 
The case of the unperturbed crystal corresponds to $R^\pr/R=1$ and is denoted by the dashed vertical line.
Each point represents the average reduced frequency of a point-confined band and the dashed red lines represent the edges of the band gap of an unperturbed crystal.
The color and shape of the symbols correspond to their values of maximum energy density $\Omega$, as described in the legend.}
\label{fig:confmapr18rpVAR}
\end{figure}

Figure~\ref{fig:confmapr18rpVAR} depicts the map of $c=3$ point-confined bands in an inverse woodpile superlattice with small regular pores of the radius $R=0.18a$, while varying the radius $R^\pr$ of the defect pore.
Once again, we observe groups of bands emerging from the top of the band gap and descending into the gap with decreasing pore radius.
A crucial observation here is that no 3D confined bands have been found for the defect pore radii $R^\pr>0.6R$, including no confined acceptor-like bands for $R^\pr>R$.
Notably, there is also a lack of bands with $\Omega>10$, which means that for $R=0.18a$ the energy concentration is in general weaker than in the case of larger pores.
There seems to be a significant preference for both the existence and the strength of 3D confinement in structures with larger regular pores.
This discovery is of considerable practical importance, since in photonic crystal fabrication it is easier to fabricate high-quality pores when they have smaller radii~\cite{Goodwin2023Nanotechnology}. 
Our research thus offers important guidelines for manufacturing inverse woodpile superlattices for confinement experiments and applications. 

\subsection{Confinement versus regular pore radius}
\begin{figure}
\centering
\includegraphics[width=\linewidth]{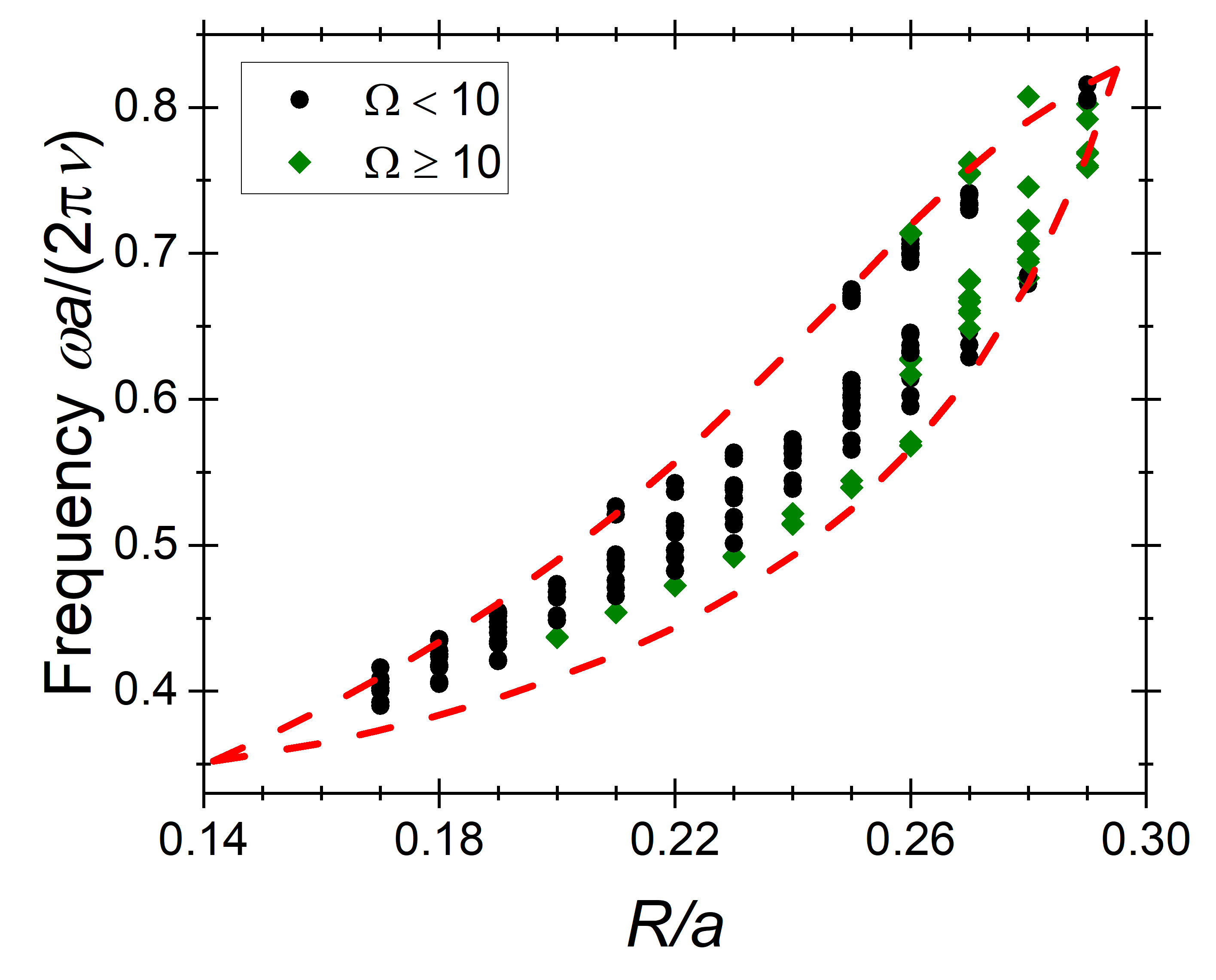}
\caption{Confinement map of point-confined $c=3$ bands in an inverse woodpile superlattice with varying regular crystal pore radius $R$ and constant ratio $R^\pr/R=0.5$. 
Each data point represents the average reduced frequency of a point-confined band and the dashed red lines represent the edges of the band gap of an unperturbed crystal.
The color and shape of the symbols correspond to their values of maximum energy density $\Omega$, as described in the legend.}
\label{fig:confmaprVARrp0.5}
\end{figure}
\begin{figure}[tb]
\centering
\includegraphics[width=\linewidth]{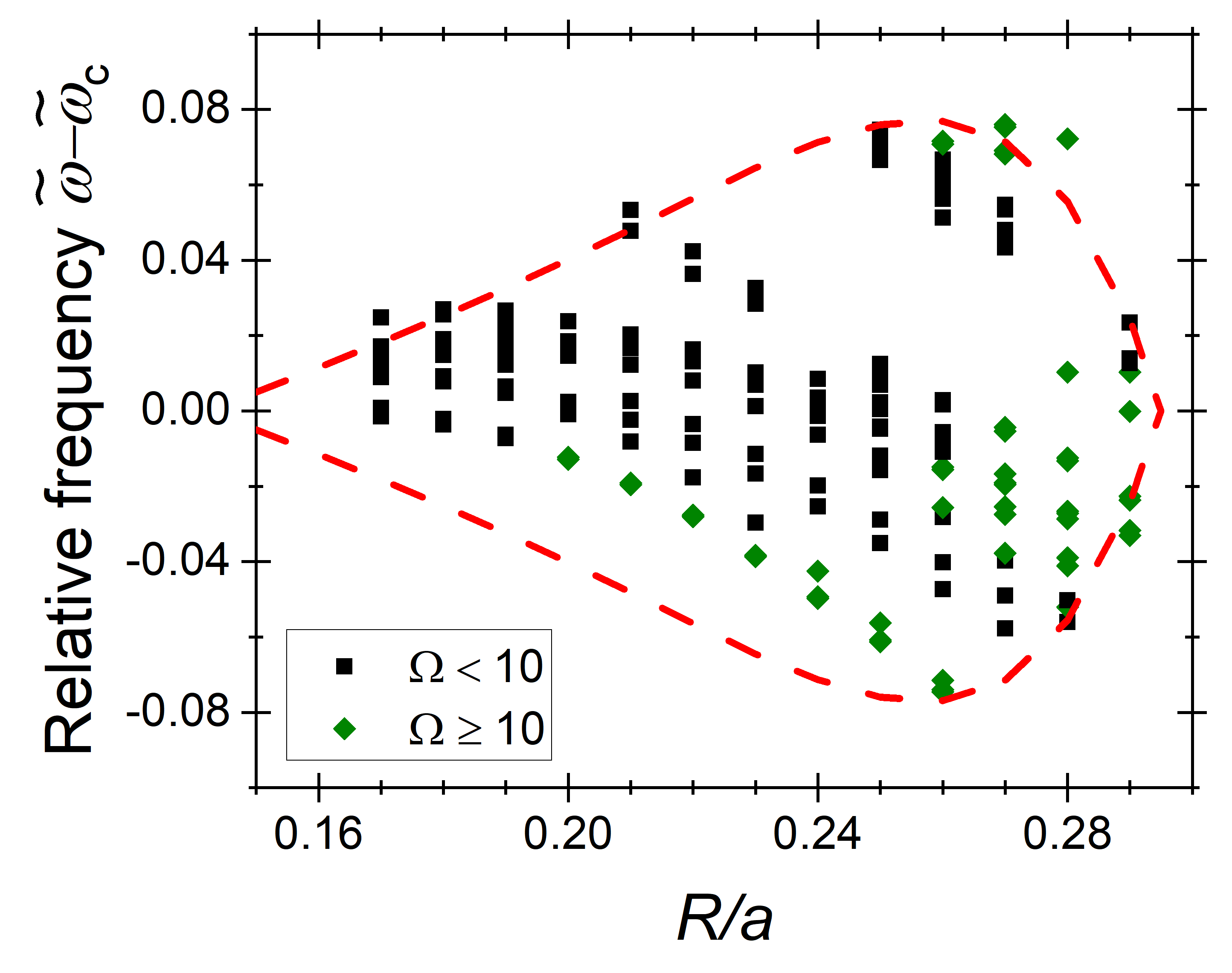}
\caption{Confinement map of point-confined $c=3$ bands in an inverse woodpile superlattice with varying regular pore radius $R$ and constant ratio $R^\pr/R=0.5$. 
This plot shows the same data as Figure~\ref{fig:confmaprVARrp0.5}, but the frequency of each structure has been adjusted by subtracting the center of the band gap for each regular pore radius $R$.
Each point represents the average relative frequency of a point-confined band and the dashed red lines represent the edges of the band gap of an unperturbed crystal.
The color and shape of the symbols correspond to their values of maximum energy density $\Omega$, as described in the legend.}
\label{fig:confmaprVARrp0.5relfreq}
\end{figure}

Figure~\ref{fig:confmaprVARrp0.5} depicts the map of $c=3$ point-confined bands in an inverse woodpile superlattice from an alternative point of view, where the ratio between the defect and the regular pore radius $R^\pr/R=0.5$ is kept constant, whereas the radius of the crystal pores $R$ is tuned. 
For viewing convenience, we also replotted this data in Figure~\ref{fig:confmaprVARrp0.5relfreq}, where we subtract the band gap center frequency $\omega_\mr{c}$ at each $R$ from the band frequencies. 

First of all, it is remarkable that even though the band gap has nonzero width, there are no 3D confined bands for crystal pore radii $R = 0.15a$ and $R = 0.16a$. 
This is not the case at the opposite side of the band gap, where 3D confined bands appear even very close to its closing at $R = 0.29a$. 
Similarly to the other studied cases, we observe that high energy densities $\Omega>10$ appear only in structures with larger pore radii $R\ge 0.20a$. 
All these observations together agree with our statement above that the strongly confined bands are more abundant in structures with larger pores. 
There is also a visible movement of the confined bands from the top of the band gap towards its bottom, as $R/a$ increases.
These bands again form groups, which get more separated in frequency from each other as the band gap becomes wider.


\section{Conclusions}
In this paper, we have performed a computational study of optical waves confined in 3D inverse woodpile photonic band gap cavity superlattices, with respect to their main structural parameters, namely the regular and the defect pore radii.
We have created maps of 3D confined bands via various cross sections through the parameter space of the two pore radii, and used these maps to analyze the influence of the superlattice structure on the confinement. 
We find that larger regular pore radii favor more confined bands and these bands also tend to have higher concentrated energy densities. 

Simultaneously, we have analyzed the symmetries of salient 3D confined bands in 3D inverse woodpile photonic band gap cavity superlattices. 
We conclude that the photonic band gap cavity superlattice bands exhibit very different symmetries compared to electronic orbitals known from solid-state physics, which is caused by the underlying crystal geometry and facilitated by the fact that our states here derive from global Bloch states, whereas atomic orbitals are localized.
We propose that attention should be given to confined band symmetries in ``photonic solid-state matter'' and their influence on the properties of these materials.

We have analyzed the potential of the inverse woodpile photonic band gap cavity superlattices for cavity QED applications.
To this end, large concentration of energy density, proportional to LDOS, must be present in the defect-pore region. 
We find that even though the acceptor-like structures with defect pores larger than the regular pores may offer higher energy concentration, this energy is mostly concentrated in small regions of silicon and decays rapidly in air.
On the other hand, the investigated donor-like structure, despite exhibiting less concentrated energy density, provides overall higher values of energy density within the cavity due to larger silicon and smaller air volumes inside. 
Therefore, donor-like structures seem to be more favorable for spontaneous emission control.

In future, more data should be gathered and analyzed to obtain even deeper understanding of the confinement behavior of inverse woodpile superlattices. 
Such a study should be extended to encompass not only 3D confined bands, but also 2D confined ones, which have been previously shown to exist in these structures as well. 
Finally, this type of investigation should be extended to other classes of photonic superlattices that are being pursued in other labs worldwide. \\

\begin{acknowledgments}
We thank Lars Corbijn van Willenswaard, Manashee Adhikary, Allard Mosk for stimulating discussions on Cartesian wave physics, Geert Brocks for helpful discussions on tight binding in atomic crystals, and Dani\"{e}l Cox for making Figure 1. 
This research is supported by the Shell-NWO/FOM programme "Computational Sciences for Energy Research" (CSER); by NWO-TTW Perspectief program P15-36 ``Free-form scattering optics" (FFSO) with ASML, Demcon, Lumileds, Schott, Signify, and TNO; by the ``Descartes-Huygens" prize of the French Academy of Sciences (boosted by JMG); and the MESA$^{+}$ Institute section Applied Nanophotonics (ANP). 
\end{acknowledgments}

%

\end{document}